# Magnetization oscillations induced by a spin-polarized current in a point-contact geometry: mode hopping and non-linear damping effects

D.V. Berkov, N.L.Gorn

*Innovent Technology Development e.V.,*
*Pruessingstr. 27B, D-07745 Jena, Germany*

**ABSTRACT**

In this paper we study magnetization excitations induced in a thin extended film by a spin-polarized *dc*-current injected through a point contact in the current-perpendicular-to-plane (CPP) geometry. Using full-scale micromagnetic simulations, we demonstrate that in addition to the oscillations of the *propagating* wave type, there exist also *two localized* oscillation modes. The first localized mode has a relatively homogeneous magnetization structure of its kernel and corresponds to the so called 'bullet' predicted analytically by Slavin and Tiberkevich (*Phys. Rev. Lett.*, **95** (2005) 237201). Magnetization pattern of the second localized mode kernel is highly inhomogeneous, leading to a much smaller power of magnetoresistance oscillations caused by this mode. We have also studied the influence of a *non-linear damping* for this system and have found the following main qualitative effects: (*i*) the appearance of frequency jumps within the existence region of the propagating wave mode and (*ii*) the narrowing of the current region where the 'bullet' mode exists, until this mode completely disappears for a sufficiently strong non-linear damping.

# I. INTRODUCTION

Since Slonczewski [1] and Berger [2] have predicted that a spin-polarized current flowing through a thin magnetic film can induce magnetization excitations (up to a complete magnetization switching) and first experimental confirmations of this effect [3] were obtained, the spin-torque induced magnetization dynamics belongs to the most intensively studied topics of the solid state magnetism [4].

Most experiments in this area have been performed in the so called columnar geometry, i.e., for a system where an electric current flows through a 'sandwich' structure consisting of two or more thin ferromagnetic layers with lateral sizes in the region ~ 50 - 200 nm separated by non-magnetic spacers. Here a substantial degree of understanding has been achieved due to high-quality experimental studies [5], detailed theoretical analysis [6] and extensive numerical simulations [7] (citations given above are by no means exhaustive).

In contrast to this situation, the interpretation of experimental results for the magnetization dynamics in the so called point-contact geometry remains controversial [8, 9, 10, 11, 12, 13]. On the one hand, this is due to a much more complicated nature of point-contact systems compared to nanopillars. In particular, the macrospin approximation, being of some help for a small nanoelement, is absolutely invalid in the point-contact setup due to a strong exchange interaction between the oscillating area under the contact with the rest of a thin film. On the other hand, the accumulated body of experimental results is much smaller than for the columnar geometry, so that a quite limited amount of data is available for a comparison with analytical theories and numerical simulations.

Even the type of spin wave excitations induced in the point-contact experiments remains a subject of an intensive discussion. In particular, experimental threshold of the magnetization oscillations onset for the *in-plane* external field is significantly lower than the value predicted by Slonczewski for the linear spin wave mode [8], as it was pointed out in [10] (when the external field is applied *normally* to the film plane and is strong enough to ensure the out-of-plane magnetization of a film, the result of Slonczewski shows a good agreement with experimental data). Another important point is that the oscillation frequency observed, e.g., in the pioneering experiment of Rippard [9] (for an in-plane applied field) is smaller than the homogeneous FMR-frequency for the thin layer with magnetic parameters reported in [9], so that an *extended* spin-wave with such a frequency could not exist. Basing on these indications, Slavin et al. suggested that the spin-wave mode excited under conditions used in [9] is a non-linear localized mode [10, 11] for which the excitation threshold and oscillation frequency turned out to be significantly smaller than for the linear propagating wave mode.

The analytical theory of Slavin et al. employs several approximations, unavoidable in such situations (see [10, 11] for details) and as such should be verified by rigorous numerical simulations. However, as we have already pointed out in [13], numerical simulations of magnetization dynamics in the point-contact setup encounter several serious methodical difficulties, what partly explains why corresponding studies are extremely rare. In [13] we have explained how these difficulties can be overcome and made an attempt to reproduce experimental results from [9] by simulating the complete trilayer system $Ni_{80}Fe_{20}$(5nm)/Co(5nm)/$Co_{90}Fe_{10}$(20nm) studied in [9] including also the Oersted field effect. We have found that the magnetization dynamics in such a system is extremely complicated. Moreover, we have observed qualitative disagreement between our simulations and experimental data. For these reasons and taking into account also that (*i*) magnetic parameters and especially the polycrystalline structure of the 'fixed' $Co_{90}Fe_{10}$ layer are not known well enough and (*ii*) reliable evaluation of the Oersted field effect without the exact knowledge of the current distribution is not possible, we decided to perform systematic studies of the simplest possible point-contact setup where the onset of the magnetization dynamics due to the spin torque can be expected: a single magnetic layer subject to a spin-polarized current within the point contact area (with the Oersted field neglected). As we shall demonstrate below (Sec. IIIA), despite these simplifications, results of our study reveal many important aspects of the spin wave dynamics in such systems (including the existence of several excitation modes) and allow a meaningful comparison both with analytical theories [10, 11] and experimental data [9].

In addition, we have studied the effect of a non-linear damping (damping depending on the magnetization change rate, as suggested in [14]), because for the large-angle spin-torque-driven magnetization oscillations such effects are expected to be especially strong. Corresponding results are presented in Sec. IIIB.

## II. SIMULATED SYSTEM AND SIMULATION METHODOLOGY

As explained above, to make the system under study as simple as possible, but still retaining the whole non-trivial physics, we have simulated the dynamics of a system consisting of a 'free' layer only with magnetic parameters corresponding to $Ni_{80}Fe_{20}$ (Permalloy) layer in the experiments described in [9]: saturation magnetization $M_S$ = 640 G, exchange stiffness constant $A = 1 \cdot 10^{-6}$ erg/cm, negligible magnetocrystalline anisotropy, layer thickness $h$ = 5 nm. For all results presented below the external field $H_{ext}$ = 1000 Oe was applied in the film plane. The $x$-axis of our coordinate system is directed along the external field, $z$-axis is the second Cartesian axis *in the film plane*, so that $y$-axis is *perpendicular* to the film. Simulated area with the in-plane size 900 x 900 nm$^2$ was discretized into $N_x$ x $N_z$ x $N_y$ = 360 x 360 x 1 cells, so that the cell size was 2.5 x 2.5 x 5 nm$^3$; periodic boundary conditions (PBC) were used. Diameter of the current flooded area $D$ = 40 nm also corresponds to the nominal point contact diameter reported by Rippard et al. [9]. It is important to note that due to such a small diameter discretization cell size larger than used by us (see above) led to inadequate discretization of this area and caused substantial artefacts by simulating the magnetization dynamics - especially when the 2$^{nd}$ localized mode with a fine spatial structure was excited.

The Oersted field of the spin-polarized current was neglected for three reasons. First, the presence of this field lead to a much more complicated magnetization dynamics and we intended to study the minimal non-trivial model to emphasize the effects of the mode localization and non-linear damping as clear as possible. Second, the proportionality coefficient between the spin-torque amplitude used in simulations and the current strength (which determines the magnitude of the external field) is not known exactly. Third, the electric current distribution in the experimental setup is also known relatively poor, adding another uncertainty to the Oersted field evaluation.

To suppress artificial interference between the 'original' spin wave emitted from the point contact area (placed in the center of the simulated square) and waves coming from PBC replica of the initial system, we have employed our method based on the spatially dependent damping coefficient (see [13] for details). In addition, we have smoothed the spatial current distribution also in the same manner as in [13] to avoid artificial generation of the magnetization pattern with the wave length twice the cell size inside the point contact area (which appear otherwise due to the abrupt jump of the current density at the border of the area below the point contact).

Magnetization dynamics excited by a spin-polarized current was simulated using basically the same software package as in [13] (which is the extension of our commercially available MicroMagus package - see [15] for implementation details). Thermal fluctuations were neglected ($T$ = 0). Spin torque acting on the magnetization **M** was included into the Landau-Lifshitz-Gilbert (LLG) equation of motion in a meanwhile standard way via the additional term $\Gamma_{st} = a_J[\mathbf{M}\times[\mathbf{M}\times\mathbf{p}]]$, where **p** denotes the spin polarization direction of electrons in the *dc*-current through the device and $a_J$ is proportional to the current strength *I*. In our simulations **p** was chosen to be *opposite* to the applied field direction $\mathbf{H}_{ext}$, because in real point-contact experiments the magnetization dynamics is supposed to be driven by spin-polarized electrons *reflected* from the 'fixed' magnetic layer (of a FM/NM/FM trilayer system) towards the 'free' one. The package was extended further in order to take into account non-linear (magnetization rate dependent) damping as suggested in [14]; details of this implementation will be presented elsewhere.

## III. SIMULATION RESULTS AND DISCUSSION

### A. Magnetization dynamics for the linear Gilbert damping

In this section we present main features of the magnetization dynamics for the standard linear Gilbert damping $\Gamma_{dis} = (\lambda / M_S)\cdot[\mathbf{M}\times(d\mathbf{M}/dt)]$, where the dissipation parameter $\lambda$ is constant; in simulations performed here $\lambda = 0.02$.

Dependence of the magnetization oscillation frequency *f* on the spin torque magnitude $a_J$ is shown in Fig. 1a with open circles; typical snapshots of magnetization configurations during these oscillations for various oscillation modes (see below) are shown in the same figure as grey-scale maps of the $m_z(\mathbf{r})$-projection (in-plane projection perpendicular to the applied field direction).

Already for the *linear* damping the system demonstrates a fairly rich magnetization dynamics. Oscillations start at the threshold value $a_{J,\,th} \approx 3.55\ (\pm 0.02)$ with the extended wave mode at the frequency $f \approx$ 14.3 GHz. This frequency is much higher than the homogeneous FMR frequency for this system ($f_{FMR} \approx$ 8.42 GHz) because already at the threshold the wave with a *large* wave vector $k \sim 1/R_c$ is exited due to the small radius $R_c$ of the point contact. When the current strength ($a_J$ value) is increased, the oscillation amplitude rapidly growths (see Fig. 1b), leading to the non-linear downward frequency shift as explained theoretically [10, 11, 20] and observed experimentally in several papers (see, e.g., papers of Kiselev et al. and Krivorotov et al. from [5]).

When $a_J$ exceeds the first critical value $a_{cr}^{(1)} \approx 4.45$, the first frequency jump occurs: transition from the extended wave mode (*W*) to the localized mode of the first type ($L_1$) takes place. (first observations of this mode in the point-contact geometry was reported by us in [12, 16]). The oscillation frequency drops below the $f_{FMR}$-value - which is an immanent feature of a localized oscillation mode - and decreases linearly and very slowly from $f(a_J = 4.5) \approx 7.45$ GHz to $f(a_J = 6.3) \approx 7.23$ GHz (see inset in Fig. 1a).

Dynamics of the transition from the *W*-mode to the $L_1$-mode for $a_J > a_{cr}^{(1)}$ is shown in Fig. 2b. Here the time-dependencies of the magnetization projection $m_x(t)$ (along the external field) averaged over the point contact area are plotted. It can be seen that after the spin-current is switched on, for $a_J$ values slightly larger than $a_{cr}^{(1)}$ the extended wave mode *W* is excited first (left inset on Fig. 2b). Then the amplitude of magnetization oscillations increases, until the magnetization *switching* takes place, so that the magnetization begins to oscillate around the direction *opposite* to the applied field. The frequency jump occurring by this switching can be seen very clearly from the comparison of the two insets on Fig. 2b.

At the second critical value $a_{\rm cr}^{(2)} \approx 6.3$ the next frequency jump accompanying the transition to the second type ($L_2$) of the localized mode occurs. Typical dynamics of this transitions is shown in Fig. 2c. The most interesting feature of this dynamics is that the formation of the second localized mode $L_2$ from the initial equilibrium magnetization state takes place via the intermediate formation of the first localized mode $L_1$. For the current values not much higher than $a_{\rm cr}^{(2)}$ the first localized mode can still exist for times much larger than its oscillation period. We shall return to discussion of this metastability below.

A snapshot of the spatial magnetization distribution within this second spatially localized mode $L_2$ is shown as the grey-scale map of $m_z(\mathbf{r})$ in Fig. 1a. This mode was found by us previously in the double-layer system [13], where it was the only type of the stable localized mode. It was shown that the core magnetization structure of this mode consists of two vortex-antivortex pairs. The frequency of the $L_2$-mode $f \approx 4.6$ GHz remains nearly constant when the current strength is increased further. We are not aware of any analytical predictions concerning this kind of a localized mode.

It is highly instructive to analyze dynamical transition between different mode types from the 'energetical' point of view.

Corresponding time dependencies of the system energy for the transition $W \to L_1$ from the extended wave mode to the first localized mode are shown in Fig. 3. Here we display time dependencies of the energy *differences* $\Delta E = E(t) - E_0$ between the energy values $E(t)$ and the energy in the initial state $E_0$. We plot these differences for the total system energy $E_{\rm tot}$ and the standard micromagnetic contributions to $E_{\rm tot}$, namely, the energy in the external field $E_{\rm ext}$ (Fig. 3b), exchange energy $E_{\rm exch}$ (Fig. 3c) and stray field energy $E_{\rm dem}$ (Fig. 3d); we remind that the small magnetocrystalline anisotropy of Permalloy was neglected, so that the anisotropy energy $E_{\rm an} = 0$. All energies are evaluated for the magnetization configuration *of the total simulation area* (i.e. *not* only for the point contact area !). Plots shown at Fig. 3b-d prove that by the transition $W \to L_1$ all partial energy contributions decrease, although both the oscillation amplitude and the magnetization gradient *within the point contact* area for the $L_1$-mode are much larger than for the $W$-mode. However, strong localization of the magnetization oscillations for the $L_1$-mode compensates for these increments, because for the extended mode $W$ the *whole* simulation area contributes to the total system energy.

The picture for the second transition $L_1 \to L_2$ is qualitatively different (see Fig. 4), because here *both* modes are localized. By this transition the exchange energy $E_{\rm exch}$ (Fig. 4c) increases - due to the steeper spatial variation of the magnetization within the point contact area for the mode $L_2$ compared to $L_1$. This exchange energy increase is compensated, first, by the decrease of the stray field (demagnetizing) energy $E_{\rm dem}$ due the formation of a magnetic charge 'quadrupole' by the two vortex-antivortex pairs characterising the $L_2$-type. The stray field energy of such a quadrupole is significantly lower than $E_{\rm dem}$ of the approximately homogeneous magnetization configuration of the $L_1$-mode kernel. Second, the average mode energy in the external field $E_{\rm ext}$ also decreases, due to a much smaller magnitude of the average magnetization $\langle m_{\rm av} \rangle$ of the mode kernel for the $L_2$-type. Again, the reason for this smaller magnitude of $\langle m_{\rm av} \rangle$ is a more complicated inhomogeneous structure of the $L_2$- kernel.

The analysis of the spatial distribution of the oscillation power within the mode kernels and the power emission of the localized modes is presented in Fig. 5. The power spectrum shown in these figures is computed by averaging the magnetization oscillation spectra obtained for each discretization cell over all cells (see [17] for details). The oscillation power maps displayed as insets for each spectral peak present the spatial in-plane distribution of the oscillation power for the $m_y(\mathbf{r})$-component at corresponding frequencies.

In accordance with the snapshots shown in Fig. 1, the spatial power distribution of the magnetization oscillations in the $L_1$-kernel shown in Fig. 5a has the elliptical symmetry with respect to the point contact center. The power dependence $P(r)$ on the distance from this center is non-monotonous; we shall return to this observation later (see discussion below). Spatial maps for spectral peaks above the basic FMR frequency give the pattern of the energy emission out of the point contact area for this mode. One can see that the angular distribution of this energy emission is highly anisotropic, whereby for different frequencies the energy is emitted in different directions. This radiation anisotropy should be explicitly taken into account in experiments and technical applications where the synchronization of the nanocontact oscillators in the in-plane geometry is aimed. In addition, our results demonstrate that numerical simulations where the spin torque effect in nanocontacts is simulated via an artificial 'localized magnetic field' (like those presented in [18]), lead to a qualitatively incorrect picture of the power emission out of the point contact area. Hence such oversimplified simulations can by no means be used for the analysis of spin-current induced magnetization excitations in such systems, not to mention the optimization of the nanocontact oscillators synchronization.

The spatial structure of the oscillation power distribution for the $L_2$-mode (see Fig. 5b) is highly complicated already for the mode kernel (leftmost inset in Fig. 5b). The energy emission pattern is also more complicated than for the $L_1$-mode (see middle and right insets in Fig. 5b); in addition, the preferred direction of the energy emission changes with the current strength $a_J$ (results not shown). We also remind that the average oscillation power at the basic frequency of this mode is much lower than the corresponding power for the $L_1$-mode (see Fig. 1b), which is also due to a more inhomogeneous magnetization configuration within the mode kernel (region under the point contact area).

We begin the discussion of our results with their comparison to known analytical theories.

_The extended wave mode_. The propagating wave mode $W$ which in our simulations is excited first when the current is increased, corresponds to the linear mode studied by Slonczewski in [8]. The threshold current $I_{th}$ for the excitation of this mode can be written for our purposes in the form proposed in [11] as

$$I_{\mathrm{th}} \approx \frac{1}{\sigma}\left(1.86\frac{D(H_0)}{R_{\mathrm{c}}^2} + \Gamma(H_0)\right) \tag{1}$$

where the common factor $\sigma$ depends on the magnetic layer thickness $d$, point contact area $S_{\mathrm{c}} = \pi R_{\mathrm{c}}^2$, material saturation magnetization $M_S$ and the spin polarization degree $P$ of the current as $\sigma = g\mu_{\mathrm{B}} P/(2|e|M_{\mathrm{S}} \cdot d \cdot S_{\mathrm{c}})$. The first term on the right-hand side of (1) describes the energy loss due to energy flow carried out of the point contact area by the extended circular (elliptical) wave. This term is proportional to the spin wave dispersion $D(H_0)$, which for the field-in-plane geometry has the form

$$D(H_0) = \frac{2\gamma A}{M_{\mathrm{S}}} \cdot \frac{H_0 + 2\pi M_{\mathrm{S}}}{\sqrt{H_0(H_0 + 4\pi M_{\mathrm{S}})}} \tag{2}$$

The 'normal' energy dissipation within the point contact area is given by the second term on the right-hand side of (1), which for the same geometry is $\Gamma(H_0) = \gamma\lambda \cdot (H_0 + 2\pi M_{\mathrm{S}})$. Taking into account that the product of the coefficient $\sigma$ and the current strength $I$ is related to our spin torque magnitude $a_{\mathrm{J}} = \hbar \cdot I \cdot P/(2|e|M_{\mathrm{S}}^2 \cdot d \cdot S_{\mathrm{c}})$ [19] via $\sigma I = \gamma M_{\mathrm{S}} \cdot a_{\mathrm{J}}$, it is easy to derive the analytical threshold value $a_{\mathrm{J,\,th}}^{\mathrm{an}}$ for the onset of the Slonczewski mode:

$$a_{J,\text{th}}^{\text{an}} = \frac{H_0 + 2\pi M_S}{M_S} \cdot \left[ \frac{1.86}{R_c^2} \cdot \frac{2A}{M_S} \cdot \frac{1}{\sqrt{H_0 (H_0 + 4\pi M_S)}} + \lambda \right] \quad (3)$$

Substituting all the values which have been used in our simulations ($H_0$ = 1000 Oe, $M_S$ = 640 G, $R_c$ = 20 nm, A = $1.0 \cdot 10^{-6}$ erg/cm, $\lambda$ = 0.02), we obtain $a_{J,\text{th}}^{\text{an}} \approx 3.94$. The good agreement of this value with the threshold obtained in simulations $a_{\text{J, th}}^{\text{num}} \approx 3.55\ (\pm 0.02)$ can be viewed as an evidence for the good quality of either the simulation software or approximations used by the derivation of (1); the latter involve mainly the neglect of the group velocity anisotropy in the field-in-plane geometry.

The localized mode $L_1$. This first type of the localized mode which existence was demonstrated by us for the first time in [12, 16] corresponds most probably to the localized non-linear 'bullet', which was predicted and thoroughly analyzed using the non-linear dynamics methods by Slavin et al. [11]. First of all, we point out that the frequencies of our mode ($f(L_1) \approx$ 7.25 - 7.45 GHz) are very close to those predicted in [11] for the same set of magnetic parameters ($f_{\text{bull}} \approx$ 7.4 - 7.8 GHz). A very important feature also is that these frequencies are definitely below the homogeneous FMR frequency $f_{\text{FMR}} \approx$ 8.42 GHz for the studied thin film.

Second, our mode $L_1$ is also localized, as it is the bullet mode of Slavin et al. Comparison of the power dependencies on the distance to the contact center $P(r)$ for the propagating wave mode $W$ and the $L_1$-mode is shown in Fig. 6. It can be seen, that although the normalized oscillation power of the $L_1$-mode near the point contact center is larger than that of the $W$-mode, for $r \gg R_c$ the oscillation power decays for the $L_1$-mode much faster than for the $W$-mode. From the left inset to Fig. 6 it can be seen that the $W$-mode power decays as $1/r$ (we note in passing that due to the Gilbert dissipation, the decay law of the $W$-mode $P(r) \sim (1/r)\exp(-r/r_{\text{dec}}^{(W)})$ contains in principle also an exponential factor; however, analytical estimates give for the corresponding decay radius the value $r_{\text{dec}}^{(W)} \sim T_{\text{osc}} v_{\text{gr}} / \lambda \sim 10^4$ nm, so that this exponent can not be seen for the simulated area size used here). In contrast to the $W$-mode, the localized bullet mode should decay exponentially with the decay radius $r_{\text{dec}}^{(L_1)} \sim R_c$ [11]. Indeed, the exponential fit for the $P(r)$-dependence of the $L_1$-mode (right inset in Fig. 6) results in the value $r_{\text{dec}}^{(L_1)} = 12(\pm 1)$ nm (we remind that $R_c$ = 20 nm).

However, there exist also important discrepancies between our simulation results and analytical predictions of Slavin et al. First of all, in simulations the localized mode discussed above is excited *after* the propagating wave mode, whereas according to the analytical theory, the excitation threshold of the 'bullet' mode should be much smaller than for the linear (Slonczewski) mode. We attribute this difference to the circumstance, that, first, we start from the homogeneous magnetization state increasing the current value, and, second, we do not include thermal fluctuations. Hence the mode which is topologically the closest one to the homogeneous state, i.e., the linear mode is excited first. Thus we suppose that if we would be able to perform simulations during a sufficiently long time including thermal fluctuations, the localized mode would be excited for $a_J$-values below the threshold for the $W$-mode. Unfortunately, such simulations are out of the time range for the state-of-the-art micromagnetic codes, in particular, because thermal fluctuations reduce the accuracy of the numerical integration method by at least one order of magnitude (thus requiring the corresponding decrease of the time step for the prescribed accuracy). This effect is pronounced especially strong in systems with small discretization cells, as it is the case for the problem under study.

A strong support for the latter argument follows from the fact that the average system energy strongly *decrease* by the dynamical transition $W \to L_1$ (see Fig. 3 and the discussion above). Hence we argue that in presence of thermal fluctuations which allow to surmount the energy barrier required to excite the localized mode this mode should be excited first (see discussion in [11] about the finite mode amplitude of the 'bullet' at its excitation threshold).

Another important difference between our simulations and analytical results of Slavin et al. is the *non*-monotonous mode profile of the $L_1$-mode immediately after this mode appears (see Fig. 6), whereby analytical calculations predict the *monotonous* power decay of the 'bullet' mode with increasing distance to the contact center. Again, this discrepancy can be easily understood assuming that the localized mode observed in our simulations corresponds to the 'bullet' mode far above its excitation threshold. Thus analytical prediction for the mode profile given in [11] for current near the excitation threshold, would be invalid for such large currents, where a more complicated mode profile should exist. We also point out that the non-monotonous mode profile observed in our simulations is the necessary precursor for the transition to the second type of the localized mode $L_2$ with a highly complicated kernel magnetization configuration.

<u>*Comparison to the experimental data*</u>. Before we proceed with the comparison of our data with experimental results, we would like to establish a relation between the current strength used in a real experiment and the value of the spin torque magnitude $a_J$ employed in our simulations. As mentioned above, in the simplest theoretical approximation the corresponding relation in Gaussian units is $a_J = \hbar \cdot I \cdot P/(2|e|M_S^2 \cdot d \cdot S_c)$, where the only unknown parameter is the current polarization degree $P$. By adopting, e.g., the value $P$ =0.3 (see corresponding estimations in [6] and taking into account the relation between current units in Gaussian system and SI, we obtain that for the geometry using here $a_J = 1$ corresponds to $I \approx 2.6$ mA.

We remind that by simulating the magnetization dynamics we have made here several approximations: neglected the interaction with the 'fixed' magnetic layer, the influence of the Oersted field and thermal fluctuations. Nevertheless, comparing our data with the results of Rippard et all [9] we can explain some important experimental findings. First of all, it is clear that Rippard and co-workers observed the localized mode of the first type ($L_1$-mode), as it was also suggested in [11]. This conclusion is based, first, on the perfect agreement of frequencies between our $L_1$-mode with $f(L_1) \approx 7.3$ GHz and the mode observed in [9] (for $H_0$ = 1000 Oe this frequency for the maximal microwave power was $f_{\exp} \approx 7.2$ GHz, see Fig. 1 in [9]); we remind that our simulations do not contain any adjustable parameters. Additional strong indication for the $L_1$-mode is, that the frequency of our simulated $L_1$-mode decays *linearly* with increasing the spin torque magnitude $a_J$ over the wide range of $a_J$-values, in a qualitative agreement with the linear decrease of the oscillation frequency measured experimentally (see inset to Fig. 1b in [9]). We point out that the frequency of the propagating wave mode decays with the current strength *non*-linearly, as it can be clearly seen from our Fig. 1a.

Another important experimental fact that we can explain in our simulations is the abrupt disappearance of the microwave signal observed in [9] when the current was increased beyond ~ 8.5 mA. Namely, according to our results, at the second critical current value $a_{cr}^{(2)}$ the $L_1$-mode decays to the second type of the localized mode $L_2$. For this $L_2$-mode the oscillation power predicted by our simulations is more than 5 times smaller than for the $L_1$-mode. Oscillations with such a small power, additionally masked by thermal noise, could probably not be detected in a real experiment.

The most important disagreement with simulations remains the absence of the propagating $W$-mode in the real experiment [9]. The reason why such a mode was not observed, could be the same as discussed above when we compared our results with analytical predictions concerning the 'bullet' mode [11]. Namely, if the excitation threshold for the localized mode is lower than that for the propagating wave mode, this mode could be experimentally excited first due to thermal fluctuations, which could assist in overcoming the energy barrier required to excite the localized mode. If this our statement is correct, it might be possible to observe the propagating wave mode in the field-in-plane setup by performing experiments at low temperature and taking special precautions against the Joule heating of the nanocontact area.

### B. Effects of a non-linear damping

Taking into account that the constant damping coefficient $\lambda$ for the standard Gilbert damping considered above leads to the unphysical decrease of the dissipated power with the growing amplitude of magnetization oscillations, Tiberkevich and Slavin [14] proposed to introduce a phenomenological dependence of this damping coefficient on the magnetization change rate $d\mathbf{M}/dt$. As shown in [14], the first non-trivial term in such a dependence should have the form $\lambda = \lambda_G(1+q_1\xi)$, where the quantity $\xi = (d\mathbf{m}/dt)^2/(4\pi\gamma M_S)^2$ includes the dependence of damping on the magnetization change rate ($\mathbf{m} = \mathbf{M}/M_S$). The value of the non-linear coefficient $q_1$ should be calculated from the microscopic theory, so in the present study we consider it as a phenomenological quantity and explore the dependence of the system properties on the value of $q_1$.

Results of our simulations of SPC-driven magnetization oscillations in the same point contact setup as described above for various non-linear coefficients $q_1$ are shown in Fig. 7 (weak non-linearity $q_1 = 1$) and Fig. 9 (strong non-linearity $q_1 = 10$ and $q_1 = 20$) in comparison to the linear damping case discussed in the previous Section.

The influence of a small non-linearity ($q_1 = 1$) on the magnetization dynamics in the extended wave mode $W$ is weak: non-linear damping slightly reduces the oscillation power and amplitude (see Fig. 7b), thus increasing - also slightly - the oscillation frequency when compared to the linear case (Fig. 7a). The $a_J$-threshold where the $W$-mode loses its stability is also getting slightly larger. However, the existence region of the localized modes changes in a qualitative way: immediately above the first critical value $a_{cr}^{(1)}$ the system exhibits a transition to a *second* type of the localized mode $L_2$, and *not* to $L_1$, as it was the case for the linear damping. For still larger currents the system oscillation mode changes to $L_1$, and by further increase of $a_J$, the transition back to $L_2$ takes place, so that the current region $a_J(L_1)$ where the localized mode of the 1st kind $L_1$ exists, is limited both from below and from above by the regions where the second localized $L_2$ mode determines the magnetization dynamics: $a_J^{min}(L_1) < a_J(L_1) < a_J^{max}(L_1)$. When the value of the non-linear coefficient $q_1$ is getting larger, the lower border of this interval $a_J^{min}(L_1)$ increases, whereby the upper border $a_J^{max}(L_1)$ decreases, so that for a sufficiently strong non-linearity $q_1$ the first type of the localized mode does not exist anymore (see Fig. 9).

For large values of the non-linear parameter $q_1$ also another qualitative effect has been observed: frequency jumps occur within one and the same mode type - extended wave mode $W$, as shown in Fig. 9a. The number and magnitude of these jumps are determined by the concrete value of $q_1$: we have found, e.g., one jump within the $W$-mode for $q_1 = 10$ and 2 such jumps for $q_1 = 20$). These jumps are, of course, accompanied by a corresponding (relatively small) jumps in the wave vector magnitude of the spin wave emitted from the contact area and by small kinks on the power dependence $P(a_J)$ (Fig. 9b).

The relatively narrow stability region of the first localized mode (which additionally narrows by increasing the non-linear parameter $q_1$) results in a natural question whether this mode is stable (at least for $T = 0$) - or it is only dynamically metastable, having the life time larger than the numerical simulation reach. Namely, every point on $f(a_J)$-dependencies in Fig. 1, 7 and 9 is the result of a simulation run which corresponds to the physical time $t_{max}$ = 25 ns, so that if the $L_1$-mode is unstable for all $a_J$-values, but for $a_J < a_{cr}^{(2)}$ its lifetime is $\tau_{L_1} > t_{max}$, then this instability would not be discovered in our simulations. To clarify this question, we have plotted the life times of the $L_1$-mode as the function of $a_J$ for $a_J > a_{cr}^{(2)}$, i.e., where the $L_1$-mode decays to $L_2$ within the accessible simulation time. The obtained dependence $\tau_{L_1}(a_J)$ is shown in Fig. 8 together with the fit $\tau_{L_1}(a_J) = A / |a_J - a_0|^{\beta}$, which assumes that the life time as the function of $a_J$ *diverges* when $a_J$ approaches some critical value $a_0$. It can be seen that this functional form fits the obtained $\tau_{L_1}(a_J)$ very good. We consider this observation as an indirect evidence that the stability region for the first localized mode really exists, although the limited number of data for $\tau_{L_1}(a_J)$ in the immediate vicinity of $a_0$ does not allow to draw the final inference. Concluding this discussion, we would like to mention that, according to our preliminary studies [13] carried out on a double-layer system, the first localized mode may lose its stability also due to the magnetodipolar interaction with the 'fixed' magnetic layer in the point-contact geometry.

## IV. CONCLUSION

In this paper we have studied systematically (using micromagnetic simulations) magnetization excitations driven by the spin-polarized current injected into a thin film via a point contact. We could show that for an external field *in the film plane* such excitations may exist in form of qualitatively different modes. The possible mode types include a *propagating* wave mode $W$ as suggested by Slonczewski [8] and two strongly *localized* modes. The kernel of the 1st type of these localized modes ($L_1$) has an approximately homogeneous magnetization state, whereby the magnetization configuration of the 2nd type mode kernel ($L_2$) is strongly inhomogeneous. The localized mode $L_1$ corresponds most probably to the spin wave 'bullet' investigated analytically by Slavin et al. [11]. Comparison with the experimental results of Rippard et al. [9] leads to the conclusion that the microwave resistance oscillations observed in [9] are caused by the localized mode of type $L_1$, whereas the disappearance of oscillations by increasing current corresponds to the transition $L_1 \rightarrow L_2$. Further, we have shown that the power emission by both types of the localized modes is strongly anisotropic, what is very important for potential applications of the point-contact devices including the synchronization of various point-contact oscillators.

We have also demonstrated, that a weak non-linearity of the damping included into the LLG-equation of motion shifts the current threshold values corresponding to the transition between different mode types. Strong non-linearity causes frequency jumps already within a propagating wave mode and leads to a complete disappearance of the first localized mode $L_1$.

ACKNOWLEDGEMENT. We greatly acknowledge many useful discussions with A. Slavin. Financial support of the Deutsche Forschungsgemeinschaft (research grant BE 2464/4-1 in frames of the Priority Program SPP 1133 "Ultrafast magnetization dynamics") is also acknowledged.

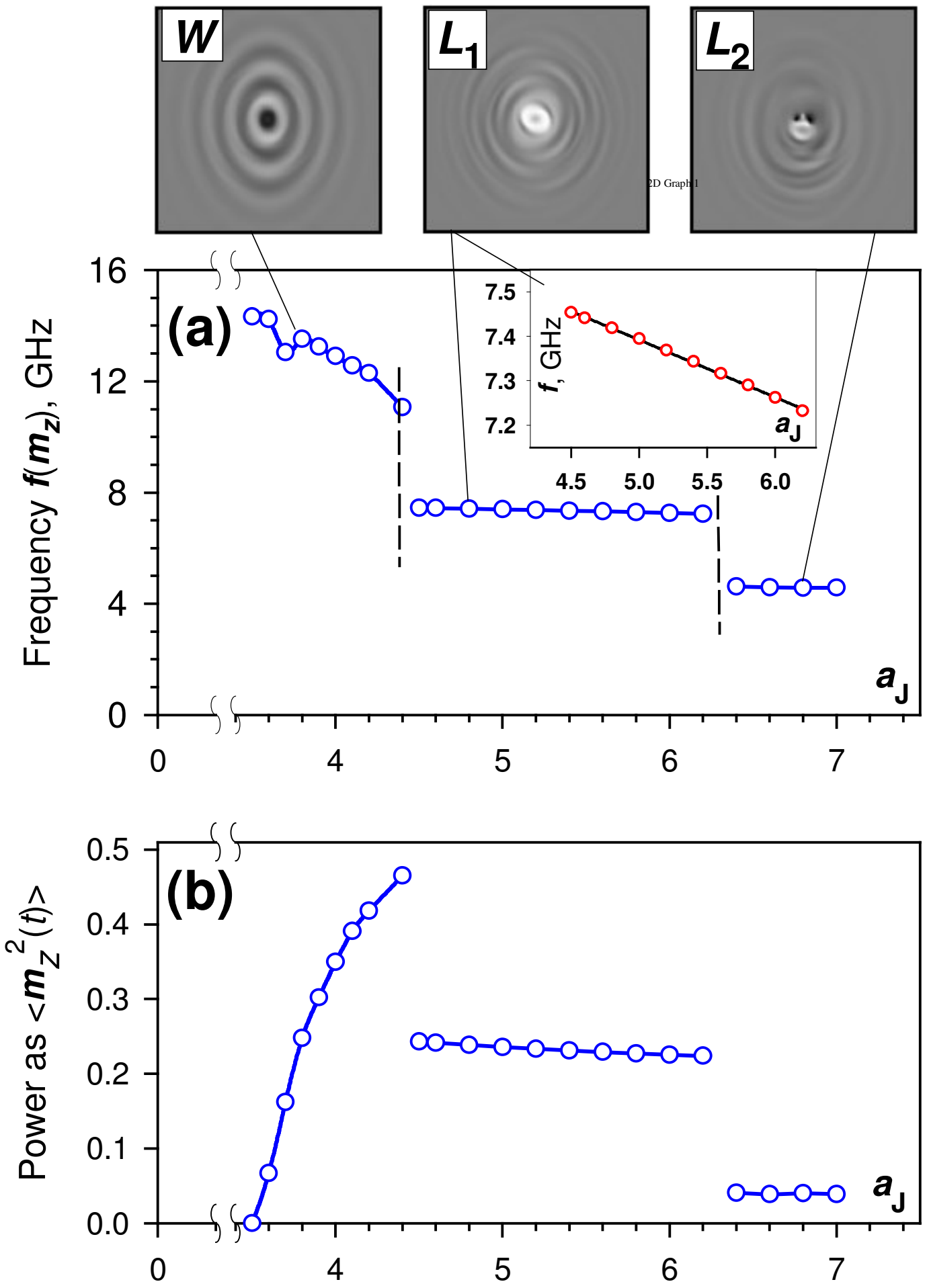

Fig.1 Dependencies of the frequency $f$ (a) and oscillation power $P$ (b) on the spin torque magnitude $a_J$ for the *linear* damping case. Frequency and power jumps accompanying the transitions from the *extended* wave mode $W$ to the *localized* mode $L_1$ and from $L_1$ to the second type of the *localized* mode $L_2$ are clearly seen. Snapshots of the $m_y$-components (perpendicular to the film plane) for all three mode types are shown at the upper panel as grey-scale maps. The inset on the panel (a) demonstrates the perfectly linear frequency dependence on the current strength for the $L_1$ -mode.

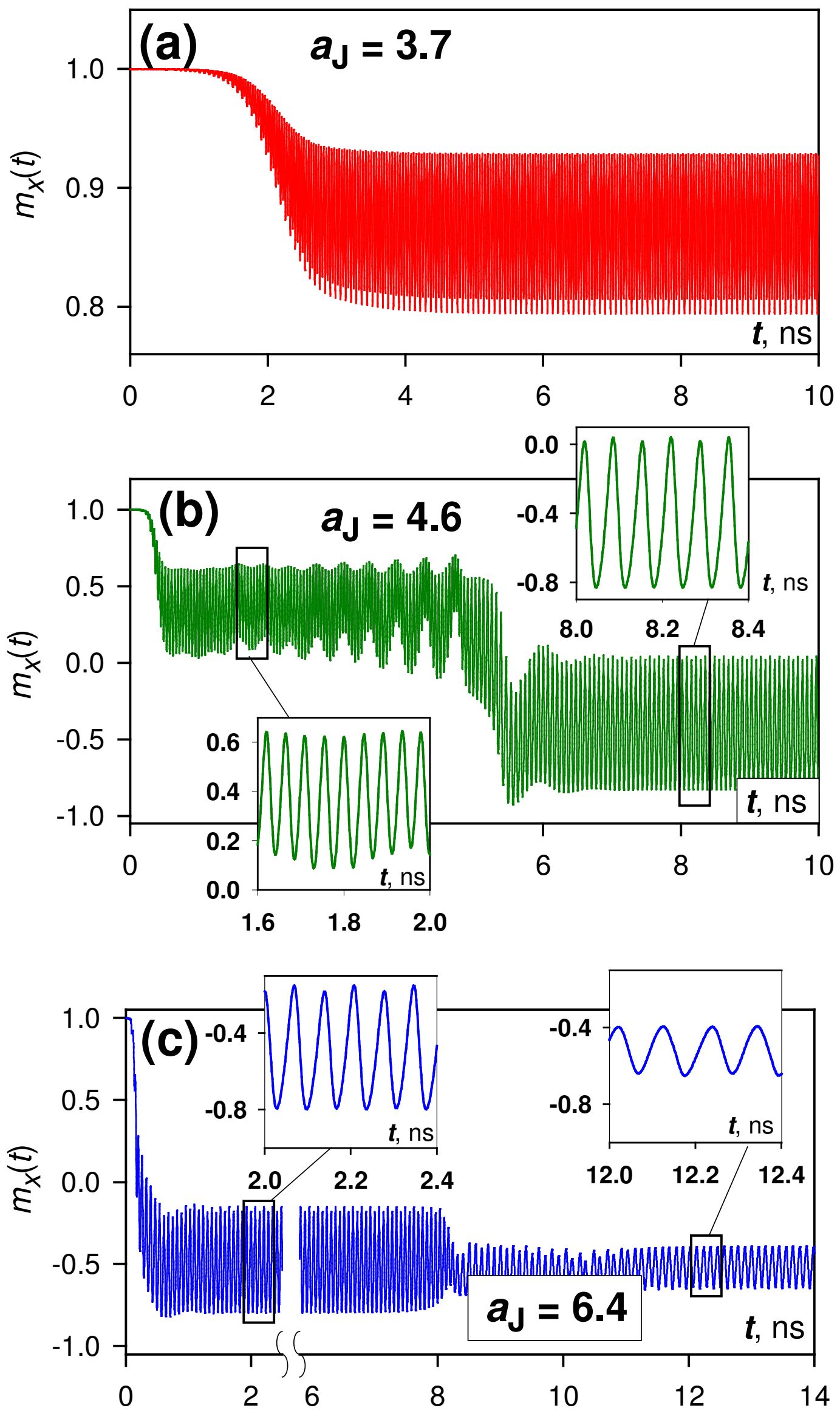

Fig. 2 Time dependencies of the $x$-magnetization component (along the external field) *averaged across the point contact area* for three different values of $a_J$: (a) - magnetization oscillations for the stable extended wave mode W for $a_J = 3.7$, (b) - the dynamical switching process leading to the formation of the localized mode $L_1$, (c) - development of the 2nd type of the localized mode $L_2$ from the $L_1$-mode. Insets demonstrate abrupt frequency jumps by the transitions between modes.

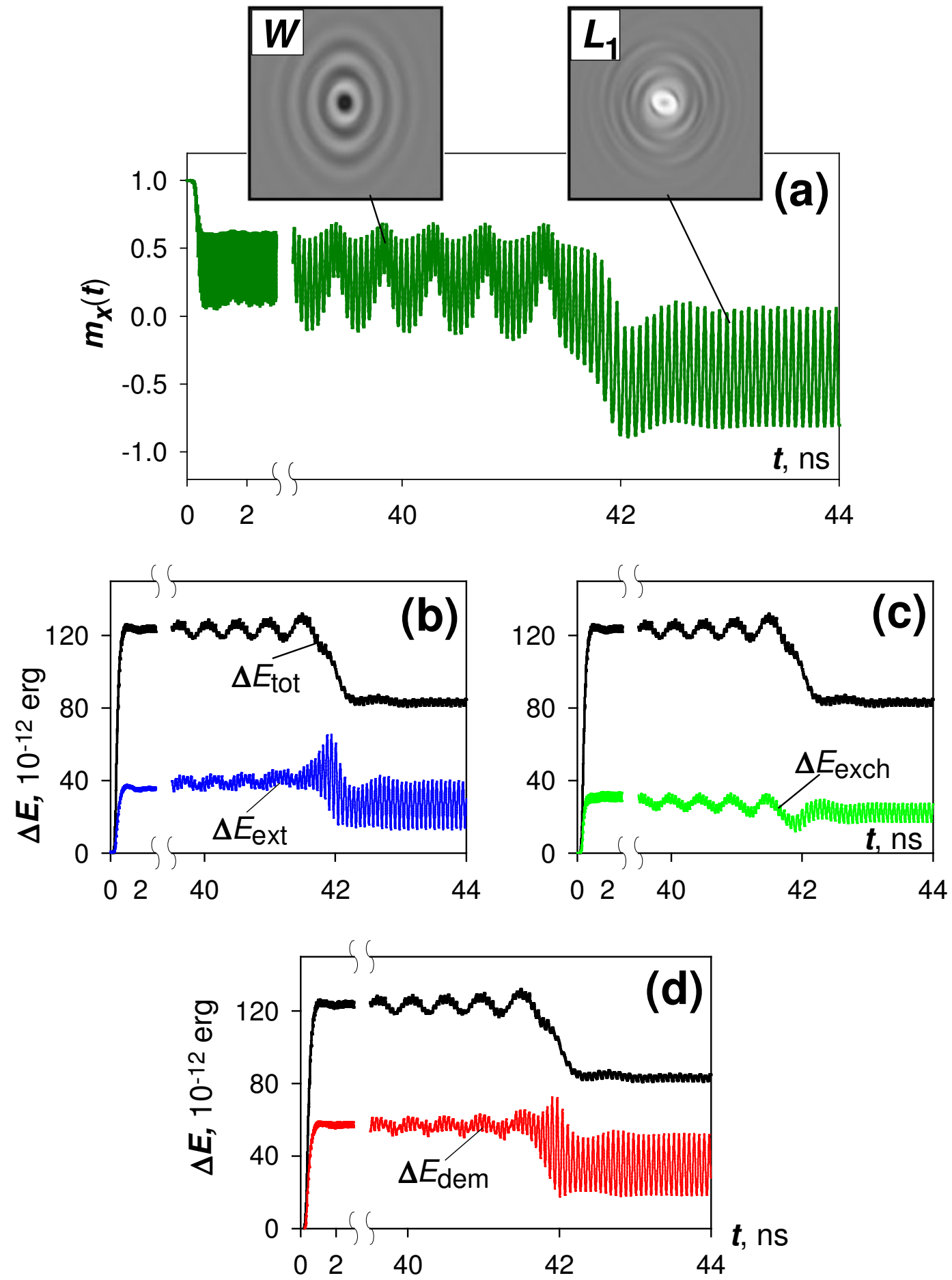

Fig 3. Energy changes of the magnetization configuration *of the total simulation area* by the transition (see panel (a))from the extended wave mode *W* to the 1st localized mode $L_1$: (b) - change of the energy due to the external field, (c) - exchange energy and (d) - demagnetizing energy. Although both the oscillation amplitude and the magnetization gradient *within the point contact* area for the $L_1$-mode are much larger than for *W*-mode, all partial contribution to the total magnetic free energy for the $L_1$-mode decrease by the transition $W \rightarrow L_1$, because magnetization oscillations in the $L_1$-mode are strongly localized.

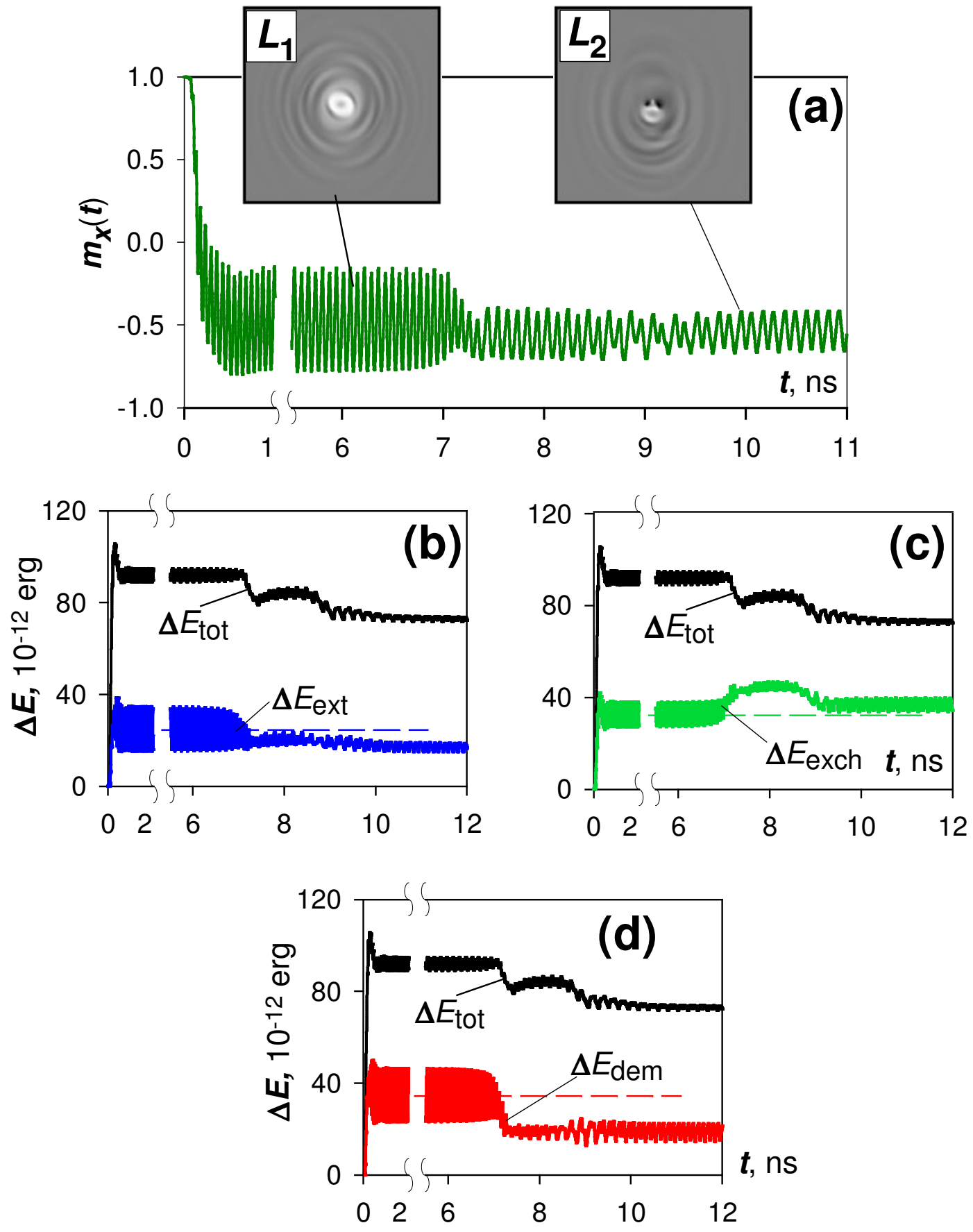

Fig. 4. The same as in Fig. 3 for the transition $L_1 \rightarrow L_2$. Here *both* modes are localized, so that the external field and the demagnetizing energies decrease due to a more inhomogeneous magnetization configuration of the $L_2$-mode in the mode localization area (resulting in the decrease of the total energy), whereby the exchange energy increases for the same reason.

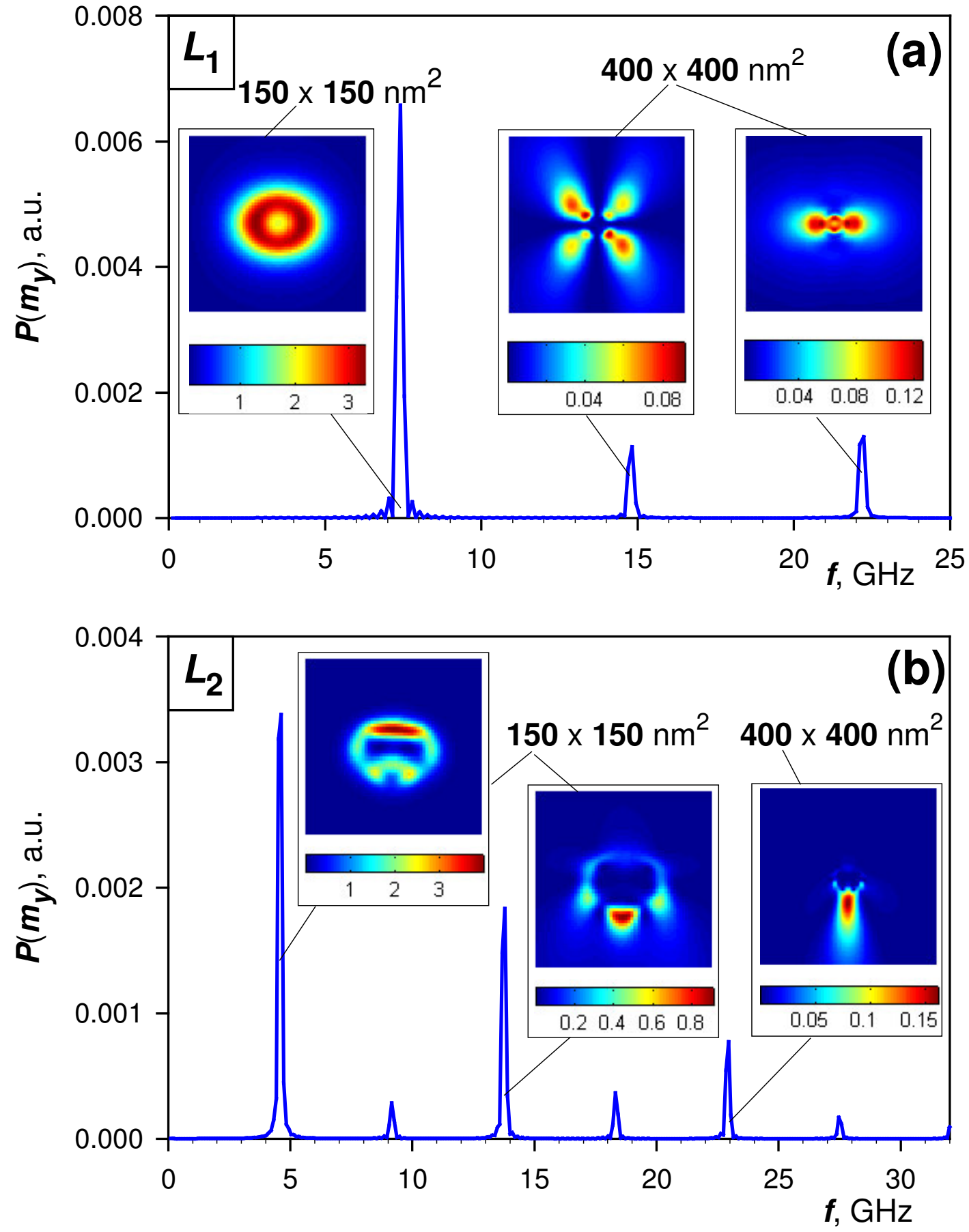

Fig. 5. Oscillation power spectra of the perpendicular magnetization component ($m_y$) averaged over the *whole* simulation area for localized modes $L_1$ (a) and $L_2$ (b). Spatial maps of the oscillation power for corresponding peaks give mode profiles (left maps) and power emission patterns out of the point contact area. Note that spatial power maps correspond to different physical sizes as written on the figure to ensure the representation of oscillations with different localization degree with an adequate resolution.

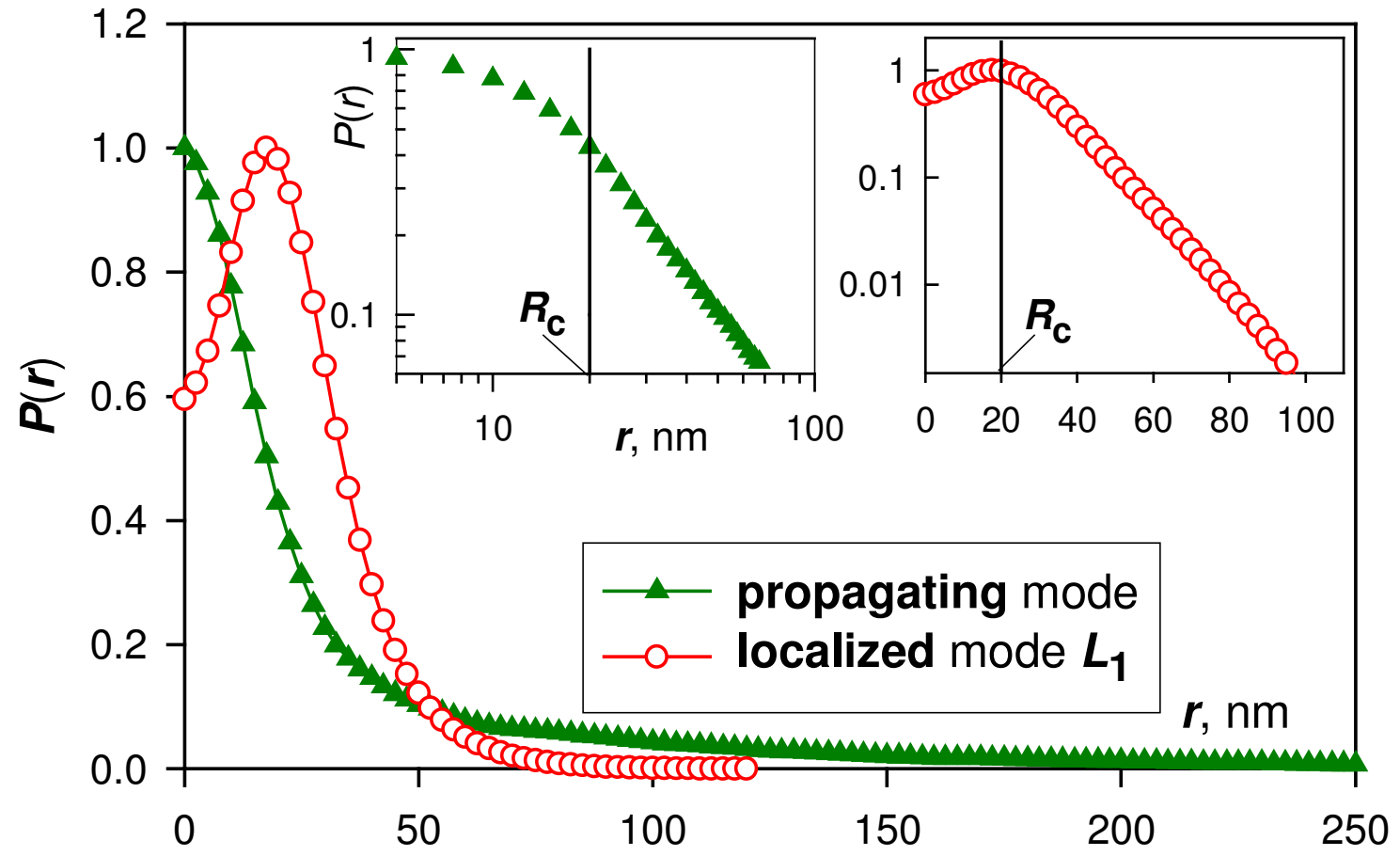

Fig. 6. Mode profiles (as the normalized magnetization oscillation power $P(r)$, $r$ being the distance to the contact center) for the propagating wave mode $W$ (triangles) and the first localized mode $L_1$. Note the non-monotonous profile of the localized mode. The left inset shows that for $r \gg R_c (= 20\text{ nm})$ the $W$-mode decays as $1/r$ (note double-log coordinates), the right inset - that the $L_1$-mode decays exponentially (note the logarithmic ordinate scale) with the decay radius $r_{\text{dec}}^{(L1)} \approx 12(\pm 1)$ nm of the same order of magnitude as $R_c$. See text for details.

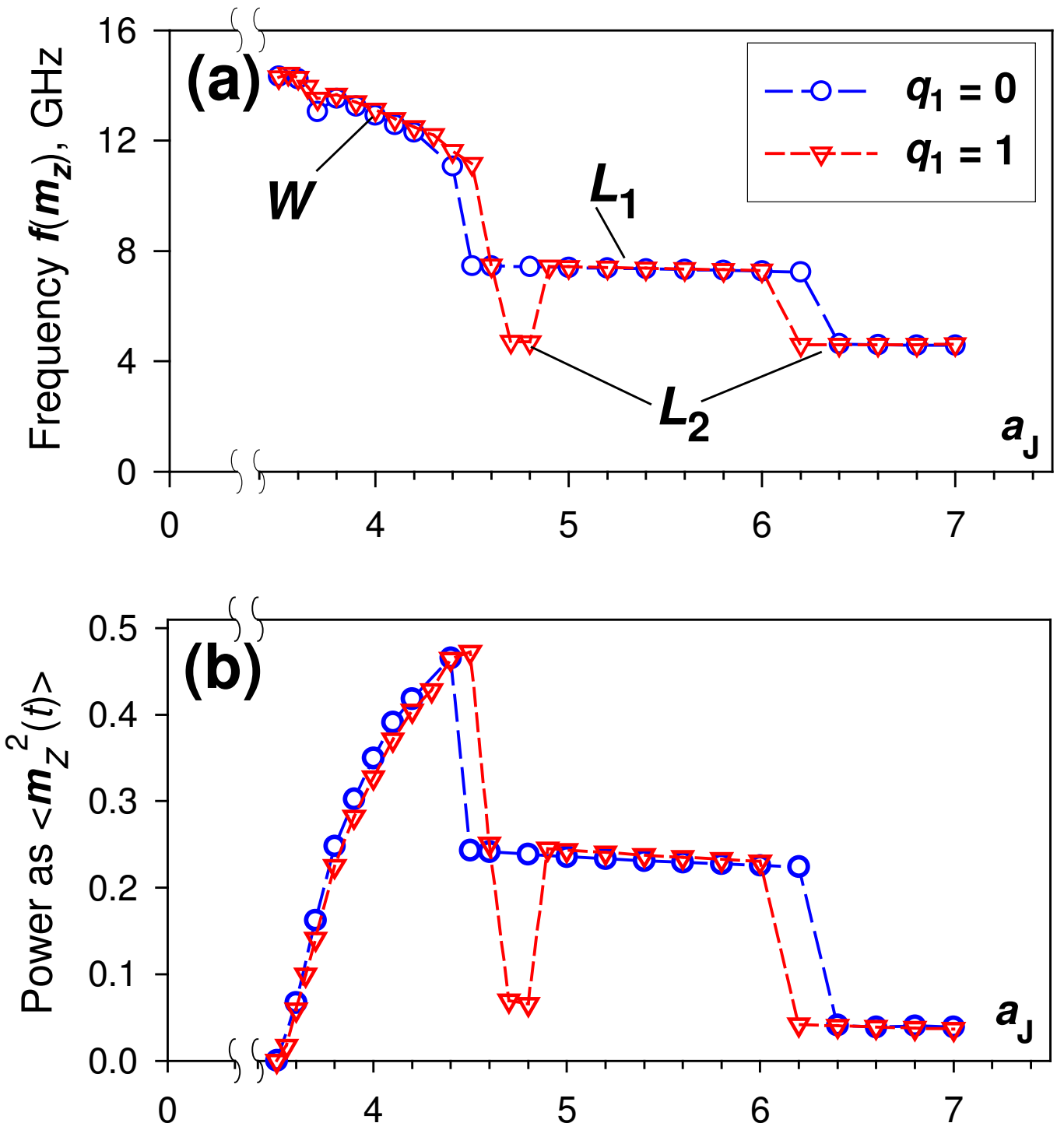

Fig.7. Dependencies of the frequency (a) and oscillation power (b) on the spin torque magnitude $a_J$ for the small non-linear damping $q_1$ = 1.0 (triangles) compared with the *linear* damping case (circles). It can be seen that for the non-linear damping the $a_J$-region where the mode $L_1$ exists is surrounded both from below and above by the existence regions of the $L_2$-mode.

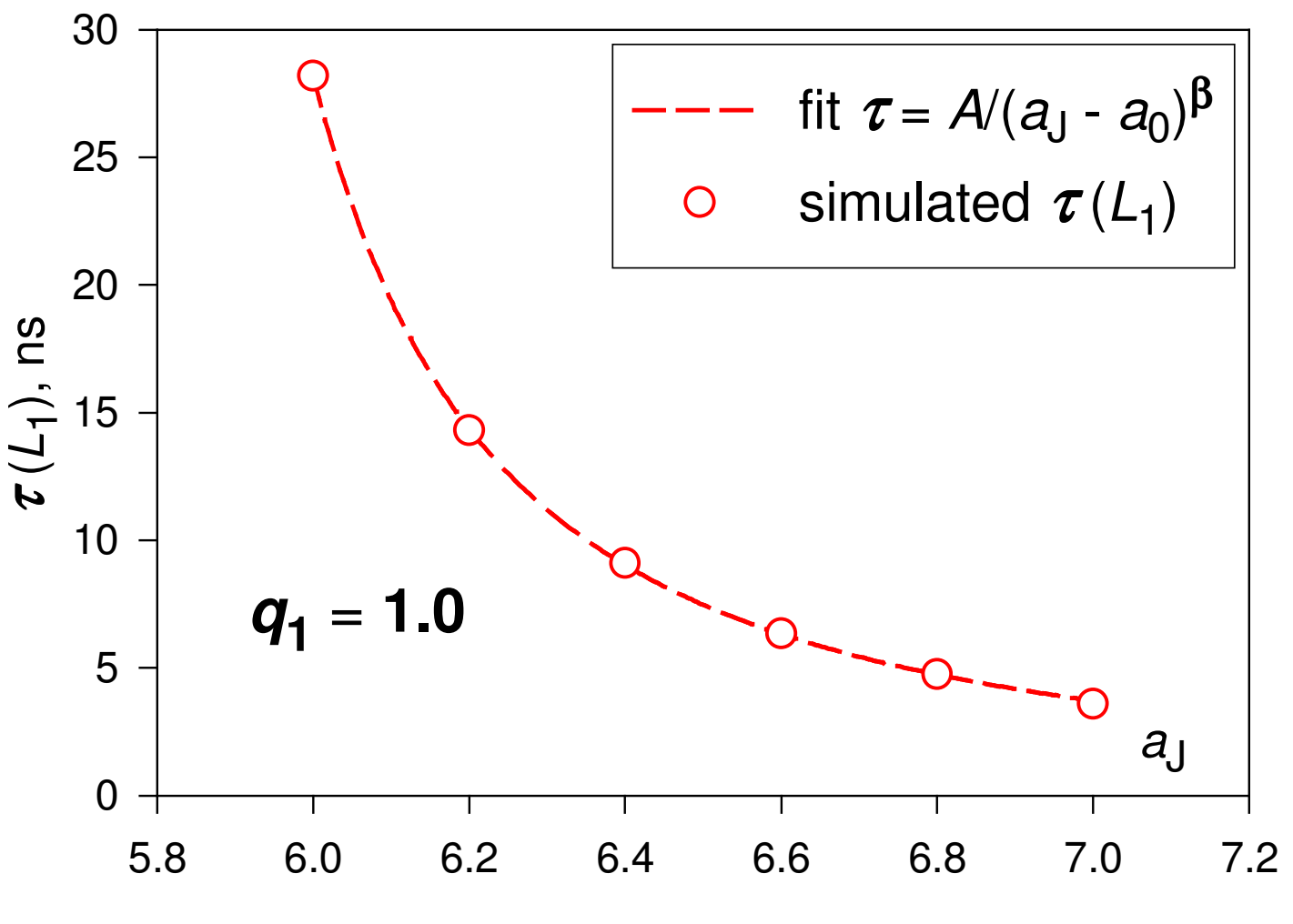

Fig. 8. Life times of the first localized mode $L_1$ for $a_J$-values where this mode decays to the $L_2$- mode. The fit of the functional form shown in the figure (with $a_0 \approx 5.6$ and $\beta \approx 1.5$) strongly indicates that this life times tends to infinity when $a_J$ decreases, so that the region where the $L_1$ -mode is stable really exists.

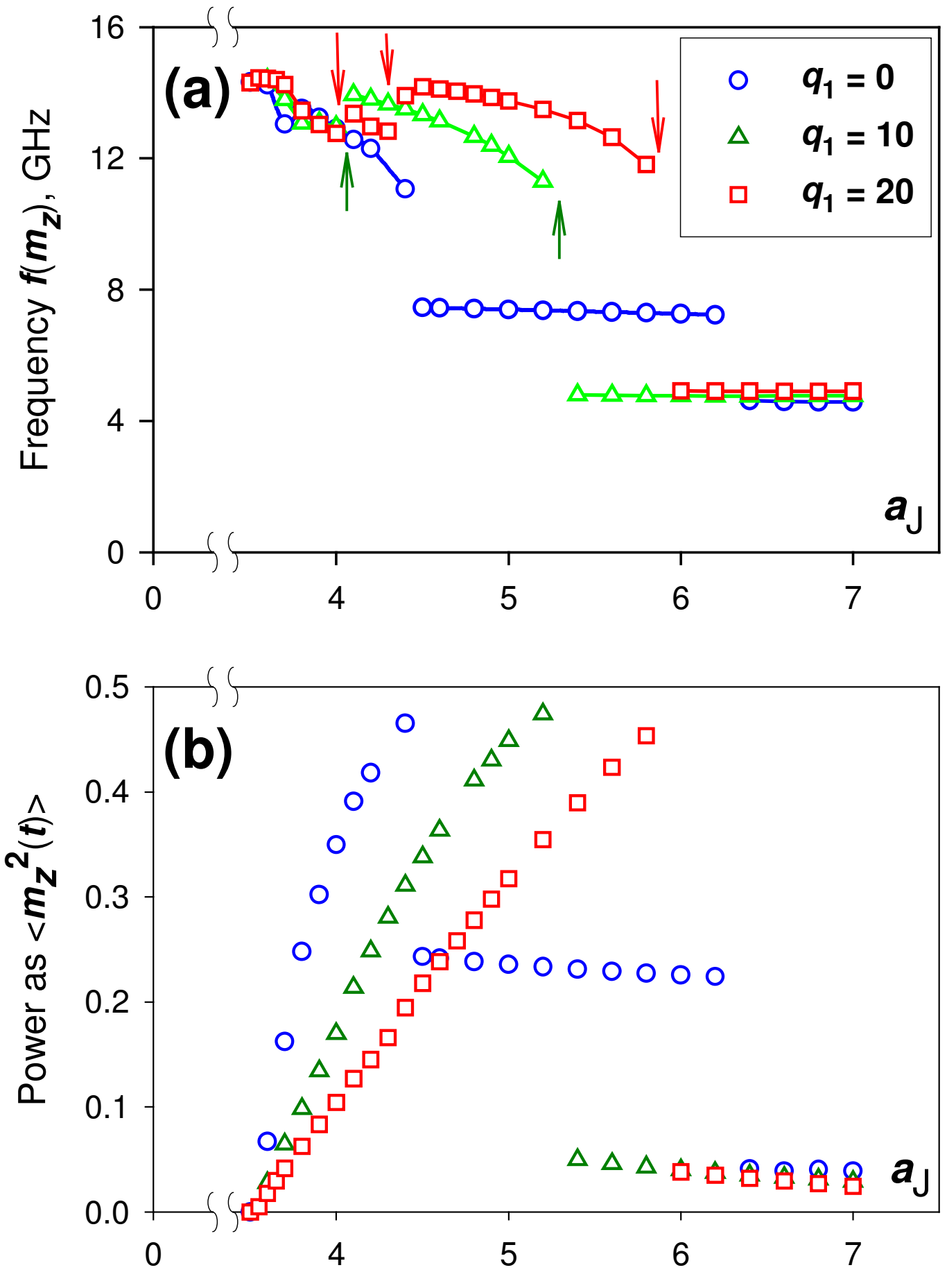

Fig. 9. The same as in Fig. 7 for the *large* non-linear damping $q_1$ = 10.0 (triangles) and $q_1$ = 20.0 (squares) compared also to the linear damping case (circles). For such large values of the non-linear parameter $q_1$ the first type of the localized mode $L_1$ does not exist anymore. Note also several frequency jumps within the existence region of the *W*-mode (shown with arrows).